\documentclass[11pt,twoside]{article}

\usepackage{asp2006}
\usepackage{lscape}

\usepackage{graphicx}
\usepackage{adassconf}

\begin{document}
\title{Spitzer's View of Edge-on Spirals}   
\author{B. W. Holwerda\altaffilmark{1}, R. S. de Jong\altaffilmark{1}, A. Seth \altaffilmark {2}, J. J. Dalcanton \altaffilmark{3}, M. Regan \altaffilmark{1}, E. Bell \altaffilmark{4} and S. Bianchi \altaffilmark{5}}

\altaffiltext{1}{Space Telescope Science Institute} 
\altaffiltext{2}{Center for Astrophysics}
\altaffiltext{3}{University of Washington}
\altaffiltext{4}{Max Planck Inst. Heidelberg}
\altaffiltext{5}{Istituto di Radioastronomia / CNR}


\begin{abstract} 
Edge-on spiral galaxies offer a unique perspective on disks. One can
accurately determine the height distribution of stars and ISM and the
line-of-sight integration allows for the study of faint structures. The
Spitzer IRAC camera is an ideal instrument to study both the ISM and
stellar structure in nearby galaxies; two of its channels trace the old
stellar disk with little extinction and the 8 micron channel is dominated
by the smallest dust grains (Polycyclic Aromatic Hydrocarbons, PAHs).
\cite{Dalcanton04} probed the link between the appearance of dust
lanes and the disk stability. In a sample of bulge-less disks they show
how in massive disks the ISM collapses into the characteristic thin dust
lane. Less massive disks are gravitationally stable and their dust
morphology is fractured. The transition occurs at 120 km/s for bulgeless
disks.
Here we report on our results of our Spitzer/IRAC survey of nearby edge-on
spirals and its first results on the NIR Tully-Fischer relation, and ISM disk stability.
\end{abstract}





\section{Tully-Fisher at 4.5 $\mu$m.}

For 32 edge-on galaxies, spanning Hubble type and mass, we fit the edge-on infinite disk model by \cite{vdKruit81a} on the IRAC mocaics; the stellar dominated 4.5 $\mu$m and the PAH emission at 8 $\mu$m, with the stellar contribution subtracted \citep[][]{Pahre04a}.
The disk's total luminosity is inferred from the fitted model: $L_{disk} = 2 \pi h^2 \mu_0$, with $h$ the scale-length and $L_0$ the face-on central surface brightness \citep[][]{Kregel02}.
%
%

%
Figure \ref{f:tf} shows the inferred Tully-Fischer relation for these disks for the stars (4.5 $\mu$m). Notably, the slope ($\alpha$) is 3.5, similar to what \cite{Meyer06a} found but contrary for the increasing trend of slope with redder filters. The effects of age and metallicity of the stellar population become independent and opposite effects on the color-M/L relation in NIR \citep[See][Fig. 2d]{BelldeJong}. 
Hence, the shallower slope in the  IRAC stellar channels, could be the metallicity effect starting to dominate.


\section{Oblateness of Stellar and PAH disks}

\begin{figure}[t]
  \begin{center}
    \begin{minipage}[t]{0.5\linewidth}
     \includegraphics[width=\textwidth]{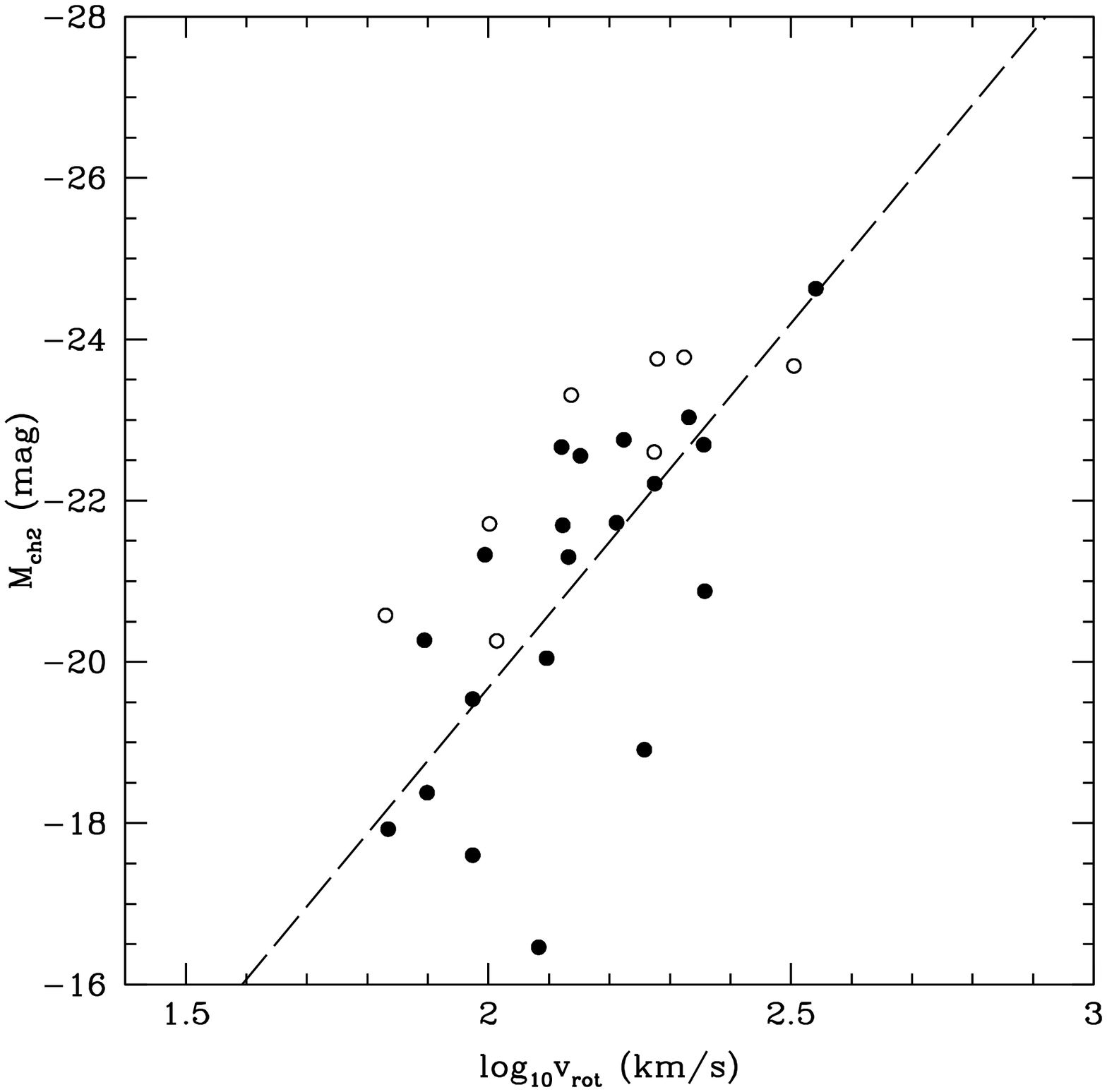}
\caption{\label{f:tf} The Tully-Fisher relation for disks in our sample. Notably, the slope is very similar to the one found by \cite{Meyer06a} for the SINGS sample. Open circles are the poorer disk fits. The slope  is probably dominated by metallicity rather than stellar age.}
    \end{minipage}\hfill
    \begin{minipage}[t]{0.5\linewidth}
  \includegraphics[width=\textwidth]{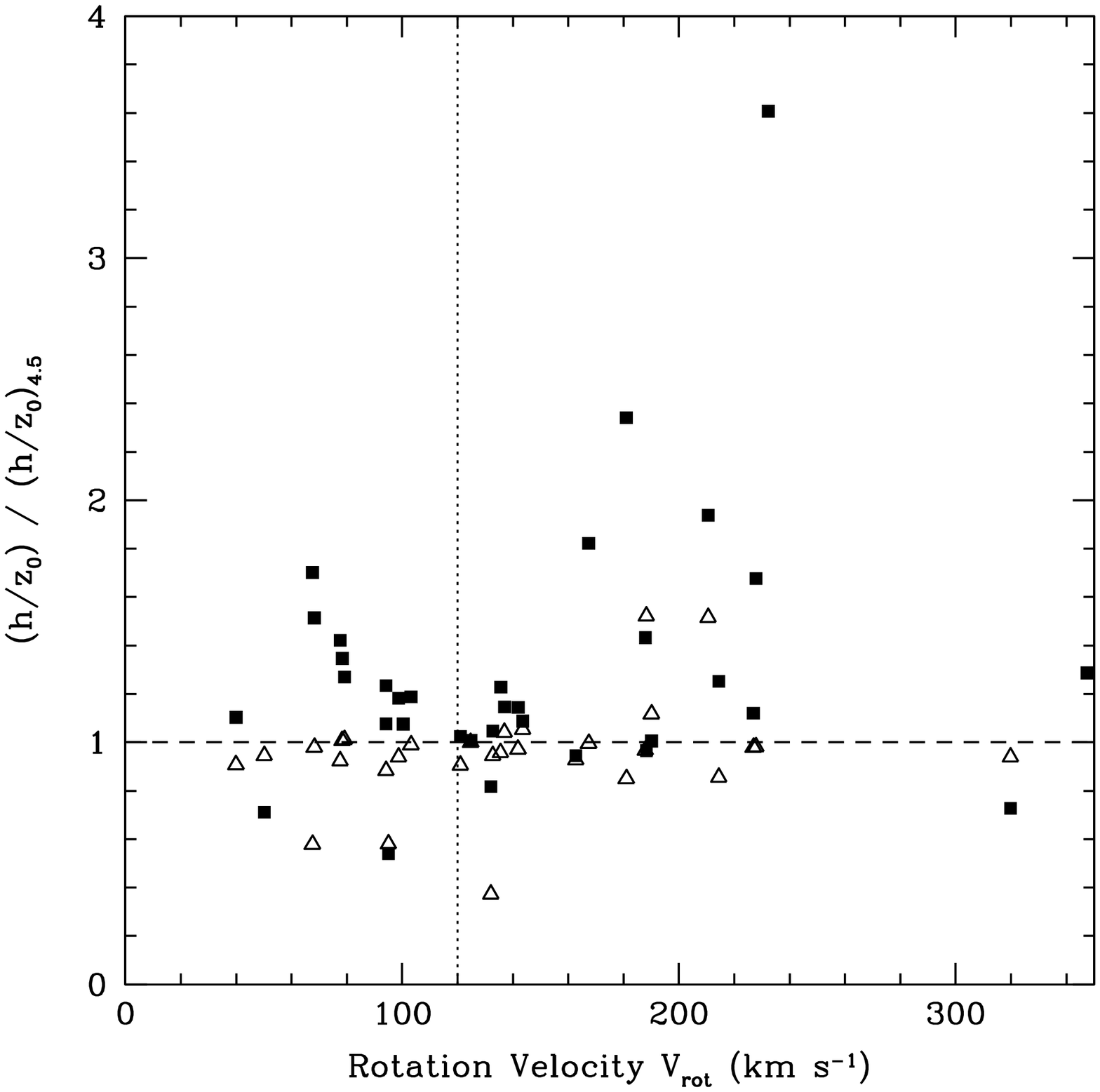}
\caption{\label{f:hz} The oblateness ($h/z_0$) of the 3.6 $\mu$m fit (open triangles) and 8.0 $\mu$m PAH map (filled squares) in terms of the 4.5 $\mu$m. The oblateness of the stellar filters, 3.6 and 4.5 $\mu$m is very similar. The PAH disk is flatter than the stellar, especially in massive disks ($v_{rot} > 120 ~ km/s$).}
    \end{minipage}
  \end{center}
\end{figure}

Figure \ref{f:hz} shows the relation between the disk oblateness ($h/z_0$) relative to the one in 4.5 $\mu$m. The 3.6 and 4.5 $\mu$m trace the stellar population and agree well. The PAH mosaic (8 $\mu$m), shows flattening compared to stellar disk, substantial in the case of the more massive disks. This appears to corroborate the reasoning of \cite{Dalcanton04} that massive disks are vertically unstable and hence form the characteristic dust lanes.

Hence, from our disk models of IRAC mosaics, we conclude: (1) The Tully-Fisher relation has a shallower slope ($\alpha=3.5$) than naively expected from trends with filter (Fig. \ref{f:tf}), and (2) The PAH disk is flatter than the stellar one, especially in more massive spirals (Fig. \ref{f:hz}).





\end{document}